 \documentclass[final,5p,times]{elsarticle}
 \biboptions{sort&compress}

\usepackage{amssymb}
\usepackage{lipsum}

\usepackage[T1]{fontenc} 
\usepackage{booktabs} 
\usepackage{ dsfont }
\usepackage{float}

\usepackage{calrsfs}
\DeclareMathAlphabet{\pazocal}{OMS}{zplm}{m}{n}
\usepackage{xcolor}
\usepackage{pict2e}
\usepackage[percent]{overpic}
\usepackage{ upgreek }
\usepackage{xfrac}
\usepackage{abraces}
\setcitestyle{numbers,square,comma,sort&compress}

\usepackage{braket}
\usepackage{stmaryrd}
\usepackage{amsmath,empheq}
 
\usepackage{multirow}

\usepackage{xurl}
\usepackage[hidelinks]{hyperref}
\usepackage{cuted}

\journal{Physics Letters B}
\begin{document}

\begin{frontmatter}



\title{Constraining meV Axion Dark Matter with ALMA Observations of the Galactic Center Magnetar SGR 1745–2900}



\author[a,b,c]{Javier De Miguel\corref{cor1}}
\author[a,b,d]{Evanthia Hatziminaoglou}
\author[a,b]{Frédéric Poidevin}
\author[e,f]{Nanda Rea}
\author[g]{Daniel L. Walker}
\author[e,f]{Davide De Grandis}
\author[b]{Jaime Prieto-Polo}


\address[a]{Instituto de Astrof\'isica de Canarias, E-38200 La Laguna, Tenerife, Spain}
\address[b]{Departamento de Astrof\'isica, Universidad de La Laguna, E-38206 La Laguna, Tenerife, Spain}
\address[c]{The Institute of Physical and Chemical Research (RIKEN), Center for Advanced Photonics, 519-1399 Aramaki-Aoba, Aoba-ku, Sendai, Miyagi 980-0845, Japan}
\address[d]{ESO, Karl-Schwarzschild-Str. 2, 85748 Garching bei M{\"u}nchen, Germany}
\address[e]{Institute of Space Sciences (ICE-CSIC), Campus UAB, C/ de Can Magrans s/n, Cerdanyola del Vall\`es (Barcelona) 08193, Spain}
\address[f]{Institut d'Estudis Espacials de Catalunya (IEEC), 08034 Barcelona, Spain}
\address[g]{UK ALMA Regional Centre Node, Jodrell Bank Centre for Astrophysics, The University of Manchester, Manchester M13 9PL, UK}


\cortext[cor1]{jdemiguel@iac.es}

\begin{abstract}
We report a mm-wave search for axion dark matter from SGR~1745--2900, based on 4.8~h of ALMA observations. No candidate features are found between 133.99--135.78, 135.91--137.70, 145.99--147.78, and 147.99--149.78~GHz, corresponding to 0.55--0.62~meV. Interpreting this null result within a state-of-the-art stellar framework, we derive sensitivity to the axion--photon coupling at the level of $g_{\gamma}\gtrsim 2\times10^{-11}$~GeV$^{-1}$ under a standard Navarro--Frenk--White profile; down to $g_{\gamma}\gtrsim 2\times10^{-13}$~GeV$^{-1}$ upon accounting for a dense dark-matter spike around Sagittarius~A*, probing the quantum chromodynamics axion parameter space.
\end{abstract}



\begin{keyword}
Dark matter \sep Axions \sep Pulsars \sep Magnetars



\end{keyword}

\end{frontmatter}




\section*{Introduction}
Axions are hypothetical pseudoscalar bosons arising from a compelling solution in quantum chromodynamics (QCD) to the strong‐interaction charge–parity problem \cite{PhysRevLett.38.1440, PhysRevLett.40.223, PhysRevLett.40.279}. Owing to their very weak coupling to ordinary matter and small mass, axions are cosmologically well-motivated candidates for dark matter (DM) \cite{ABBOTT1983133, DINE1983137, PRESKILL1983127}—an invisible component inferred from astronomical observations decades ago but still undetected at the particle level \cite{ 1933AcHPh...6..110Z, 1970ApJ...159..379R, 2018RvMP...90d5002B}. The QCD axion and axion-like particles—similar to axions but with independent mass and coupling scales \cite{PhysRevD.52.1755}—mix with photons in a static magnetic field via the Primakoff effect \cite{Primakoff:1951iae}. Classically, the axion–photon interaction relevant for this work reads
\begin{equation}
\pazocal{L} \supset g_{\gamma} \, \mathrm{\bold{E}}\cdot\mathrm{\bold{B}} \, a \,,
\label{Eq.1}
\end{equation}
where $g_{\gamma}$ is the axion–photon coupling factor, $\bold{E}$ is the electromagnetic wave and $\bold{B}$ the external static magnetic field, while $a$ is the axion.

Magnetars are highly magnetized neutron stars (NSs)—compact stellar remnants that host the most powerful magnetic fields known in nature. When DM axions enter the NS magnetosphere, they can be resonantly converted into photons, producing an axion-induced emission feature superimposed on the intrinsic spectrum of the star \cite{Iwamoto:1984ir, PhysRevD.37.1237, Morris:1984iz, Yoshimura:1987ma}. For non-relativistic axions composing cold DM, this feature is expected to lie in the radio band of the electromagnetic spectrum \cite{Lai:2006af, PhysRevLett.121.241102, Huang:2018lxq, Leroy:2019ghm, Safdi:2018oeu, PhysRevD.102.023504, Witte:2021arp, Battye:2021xvt, Darling:2020plz, Darling:2020uyo, DeMiguel:2021pfe, PhysRevD.106.L041302}.
\begin{table}[b]
\centering
\caption{Per-SPW rms in the ON and nodding (ON$-\langle$OFF$\rangle$) data
after uv-plane modelling and subtraction of Sgr~A$^\ast$. All SPWs share the same
correlator configuration and imaging parameters: $\Delta\nu_{\rm obs}=1788.9$ MHz,
$\delta\nu_{\rm ch}=7.811$ MHz, synthesized beam FWHM $0.89\times0.73$ arcsec
(PA $=-88.1^\circ$), observations carried out between MJD 60683.0 and 60700.0, and total on-source
integration time $t_{\rm int}=4.838$ h.}
\begin{tabular}{c c cc}
\hline\hline
SPW & Frequency [GHz] & \multicolumn{2}{c}{rms [mJy/beam]} \\ \cline{3-4}
    &                 & ON & ON$-\langle$OFF$\rangle$ \\
\hline
1 & 133.99--135.78 & 0.183 & 0.076 \\
2 & 135.91--137.70  & 0.151 & 0.064 \\
3 & 145.99--147.78 & 0.112 & 0.050 \\
4 & 147.99--149.78 & 0.133 & 0.062 \\
\hline\hline
\end{tabular}
\label{TableI}
\end{table}

The expected signal from ambient axions falling into the star and oscillating into photons scales with the DM density at the source location and with the magnetic field strength at the conversion surface. As discussed in Ref.~\cite{McMillan_2016}, the innermost region of the Galaxy may depart from a standard Navarro--Frenk--White (NFW) profile \cite{Navarro:1995iw} and is conjectured to host axion densities several orders of magnitude larger than in the local halo \cite{2018A&A...619A..46L, Darling:2020uyo, Darling:2020plz}. In this context SGR~1745--2900 (RA, Dec 17:45:40.164, $-29$:00:29.818 \cite{Mori:2013yda}), with a surface magnetic field above $10^{14}$\,G, emerges as one of the most favorable Galactic laboratories to probe axion DM. These advantages largely offset the observational challenges posed by its environment; the magnetar lies only $\sim$0.1\,pc from the brightest Galactic Center (GC) radio source Sagittarius~A* (Sgr~A*) and at a distance of about 8\,kpc from Earth, which implies the $d^{-2}$ geometric dilution of the signal. In this work we present the first dedicated search for axion DM using SGR~1745--2900 at millimeter wavelengths, aiming to identify narrow emission features arising from axion–photon conversion. The remainder of this Letter is organized as follows. We first describe the observations, data reduction, and analysis procedure. We then employ state-of-the-art models to estimate the sensitivity to axion DM achieved with these data. Finally, we discuss the non-detection, its implications for axion models, and our overall conclusions. 
\begin{table*}[t]
\centering
\begingroup
\setlength\tabcolsep{5pt}\small
\caption{Data analysis and matched-filter diagnostics by SPW; columns give the SPW index, fraction of channels masked (Masked frac.), fraction of channels accepted for analysis (Acceptance), median $1\sigma$ amplitude uncertainty in $\upmu$Jy ($\sigma_A$ median), mean matched-filter signal-to-noise ratio ($\langle{\rm SNR}\rangle$), standard deviation of the SNR distribution ($\sigma({\rm SNR})$), and the number of positive excursions above $3\sigma$ and $3.5\sigma$ within the SPW ($N_{\ge 3\sigma}/N_{\ge 3.5\sigma}$). Bins with $\ge 3\sigma$ are 3.14$\sigma$ at 134.016363 GHz and
3.41$\sigma$ at 134.024175 GHz. These correspond to two consecutive bins 
($\Delta\nu \approx 7.8~\mathrm{MHz}$), much narrower than the expected 
axion line width ($\mathrm{FWHM}\sim 100~\mathrm{MHz}$, spanning several 
channels).}
\begin{tabular*}{\linewidth}{@{\extracolsep{\fill}}lcccccc}
\hline\hline
SPW & Masked frac. [\%] & Acceptance [\%] & $\sigma_A$ median [$\upmu$Jy] & $\langle{\rm SNR}\rangle$ & $\sigma[{\rm SNR}]$ & $N_{\ge 3\sigma}$ / $N_{\ge 3.5\sigma}$ \\
\hline
1 & 19.6 & 77.8 & 45.36 & 0.030 & 1.208 & 2 / 0 \\
2 & 14.8 & 82.6 & 39.99 & 0.039 & 0.982 & 0 / 0 \\
3 & 23.9 & 73.5 & 35.35 & 0.038 & 0.816 & 0 / 0 \\
4 & 12.6 & 84.8 & 36.32 & -0.025 & 0.966 & 0 / 0 \\
\hline\hline
\end{tabular*}
\label{TableII}
\par\smallskip
\endgroup
\end{table*}

\section*{Observations, Data Reduction, and Analysis}
\begin{figure}[h]\centering
\includegraphics[width=.5\textwidth]{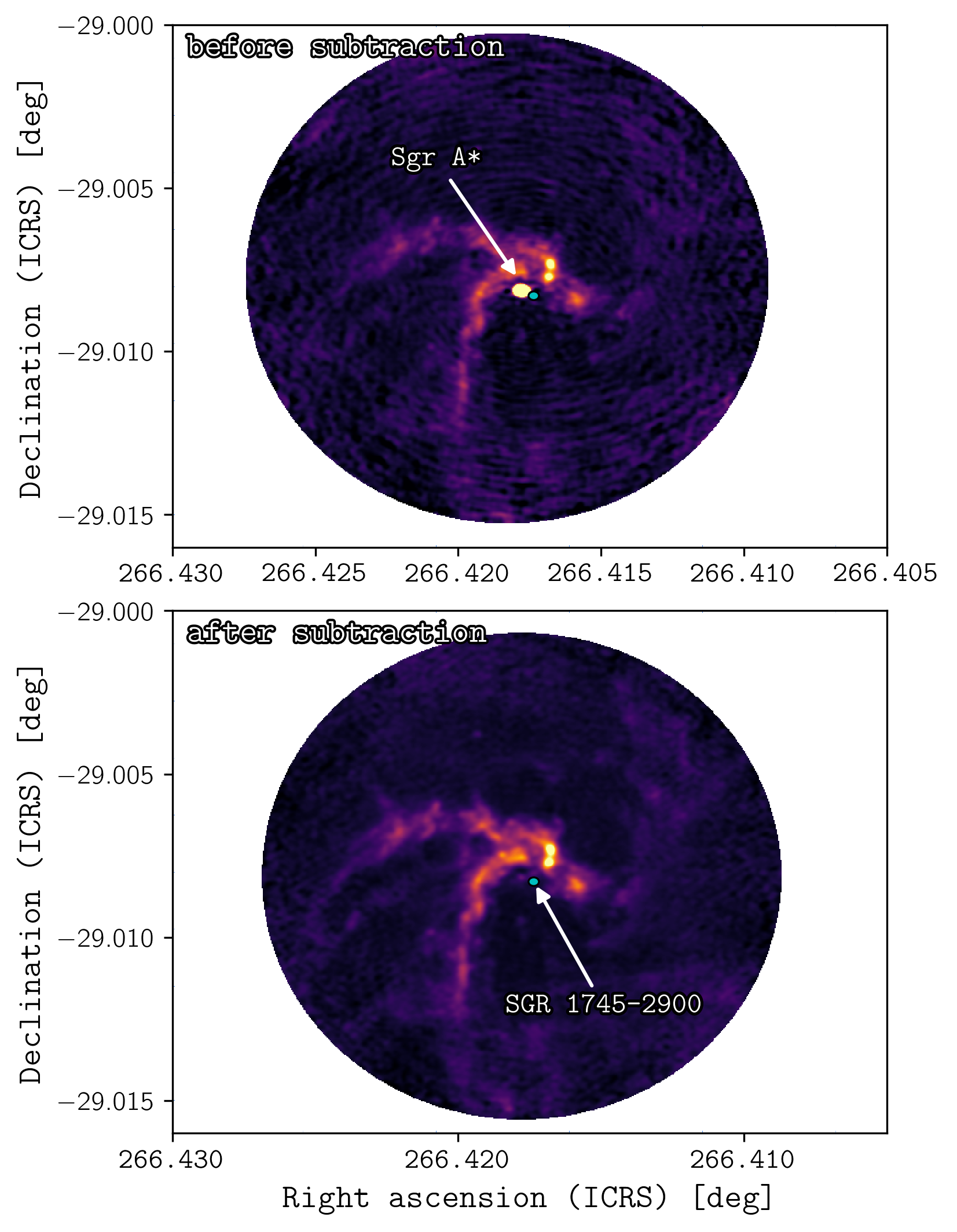}
\caption{ALMA 2-mm continuum of the GC before (top) and after (bottom) Sgr A* subtraction; the cyan ellipse marks the synthesized beam at the astrometry position of SGR 1745--2900.}
\label{Fig_0}
\end{figure}
Observations were performed with the Atacama Large Millimeter/submillimeter Array (ALMA) in the 12-m setup with an angular resolution of 0.724 arcsec and a field of view of 41.039 arcsec centered on SGR 1745--2900, with a spectral resolution of 7.8113 MHz. A total of four spectral windows (SPWs) centered at 134.9, 136.8, 146.9, and 148.9\,GHz, each with a bandwidth of $\approx$$1.8$\,GHz in ALMA band 4 were observed for a total of 4.838 h effective integration time between 2025-01-08 and 2025-01-25.

Interferometric data were reduced as follows.
All data reduction was performed with \textsc{CASA} (the Common Astronomy Software Applications, v6.6.1.17) \cite{CASATeam:2022zxb} following standard ALMA procedures. After initial flagging of corrupted visibilities, we applied flux, delay, bandpass, and gain calibrations provided by the observatory. Atmospheric transmission corrections were included through the water vapor radiometer data. We then carried out phase self-calibration using the bright continuum of Sgr A*, which improved coherence across all spectral windows \cite{albentosaruiz2025novelcalibrationimagingmethod}. The continuum emission was modeled and subtracted in the uv (Fourier) domain to minimize contamination, and spectral line image cubes were produced with pixel scales oversampling the synthesized beam. Residual narrow-band spectral features in the Sgr~A* continuum, attributable to known molecular transitions—e.g., $^{13}$CH$_3$OH, H(36)$\alpha$, HC$_3$N, H$_2$CS, SO, $^{13}$CS, CH$_3$NC, D$_2$CS, D$^{13}$CO, C$_4$H, DC$_3$N, CS, cis-CH$_2$OHCHO—were cleaned during deconvolution. At the location of SGR 1745–2900, spectra were extracted by summing over the synthesized beam in each channel, and corrected for frequency-dependent beam size. The resulting spectra typically reach the theoretical noise level expected for the achieved integration time and channel width—see Table \ref{TableI}. 
This reduction procedure ensures that the resulting cubes are optimised for isolating the spectrum of SGR~1745$-$2900 against the bright, variable supermassive black hole (SMBH) background—see Fig. \ref{Fig_0}.

\begin{figure*}[h]\centering
\includegraphics[width=1\textwidth]{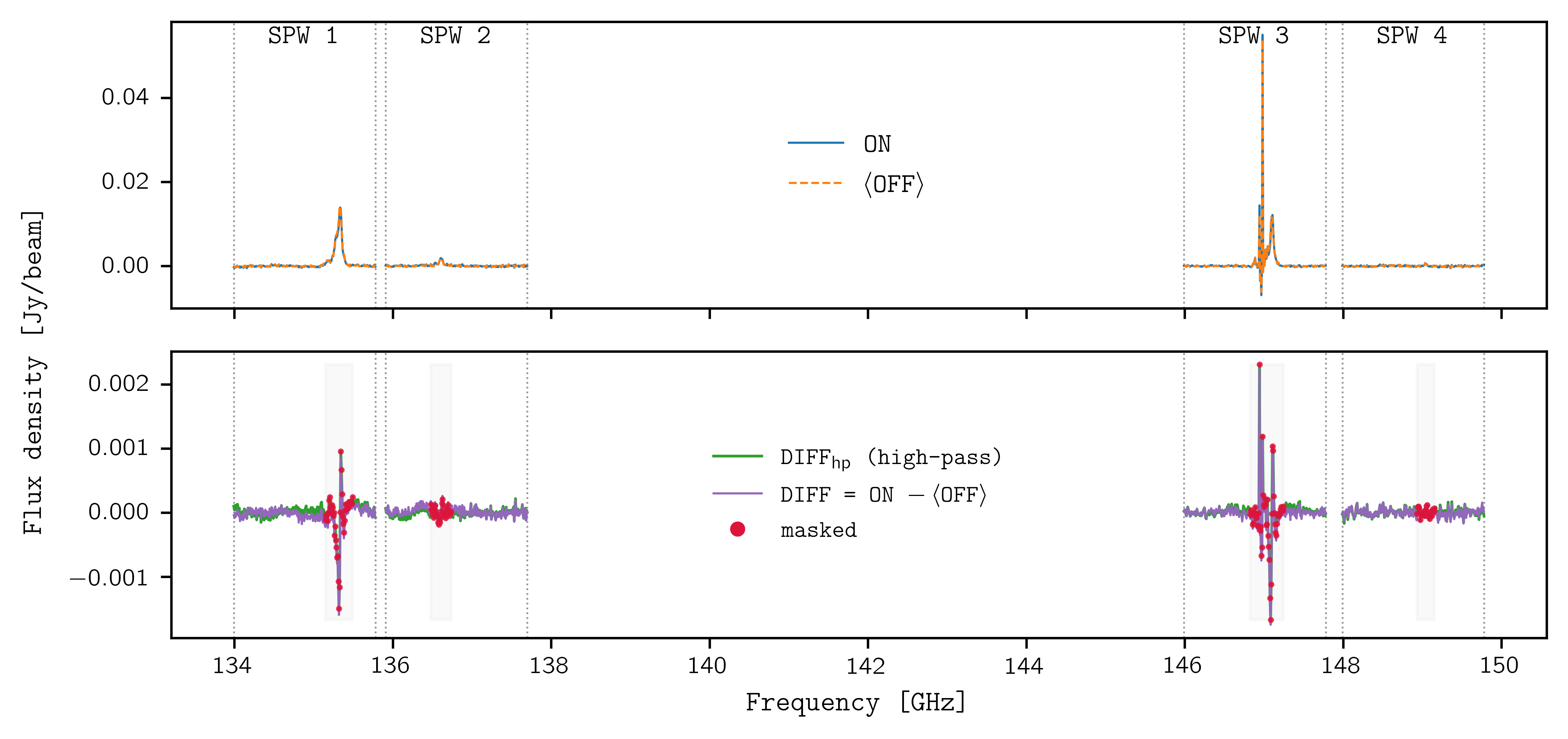}
\caption{Top: raw ON–source spectrum toward SGR~1745–2900 (blue) and averaged OFF reference (orange), shown per spectral window (SPW; vertical dotted lines).
Bottom: differenced spectrum, ON–$\langle$OFF$\rangle$ (purple), and its high–pass (hp) filtered version (green). Masked channels are marked in red with shaded regions indicating their extent.}
\label{Fig_1}
\end{figure*}

\begin{figure}[h]\centering
\includegraphics[width=.5\textwidth]{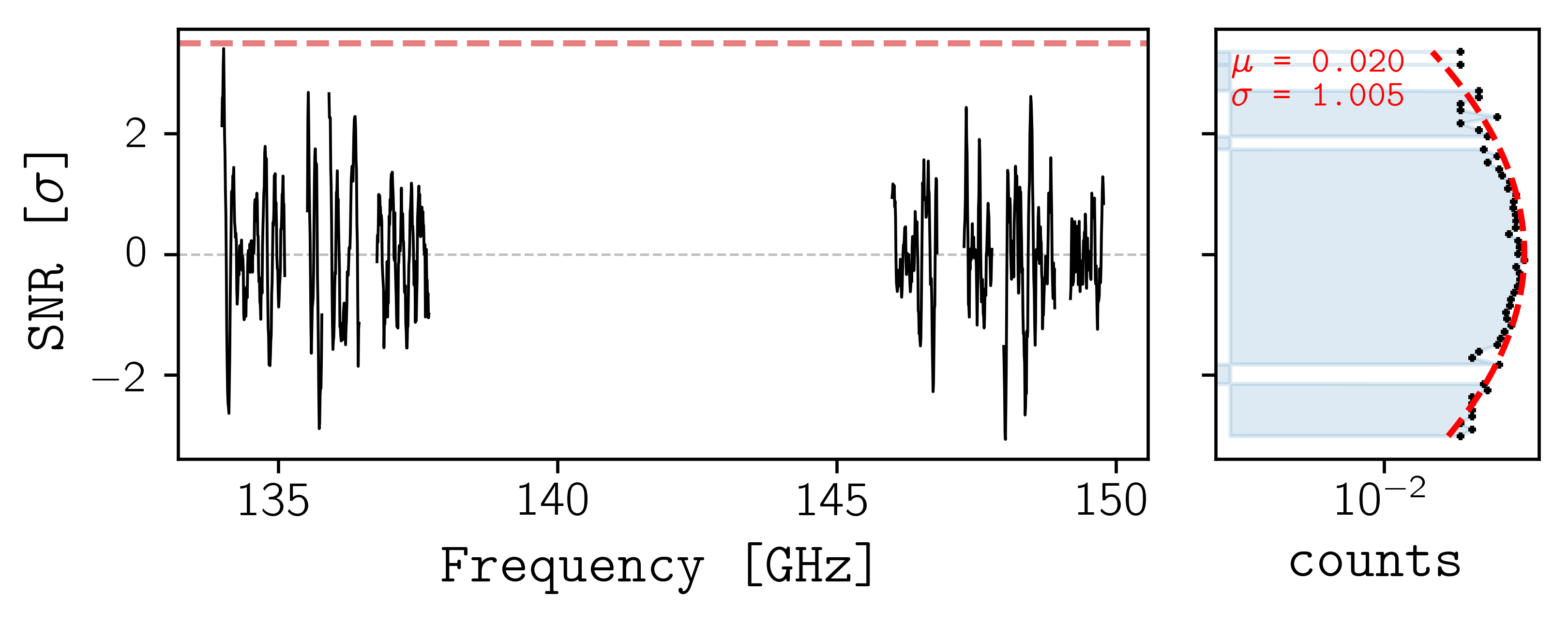}
\caption{Left: distribution of SNR across frequency after applying the matched–filter pipeline. Horizontal dashed lines indicate the zero level and the $3.5\sigma$ detection threshold (red). Right: histogram of SNR values for all accepted channels (black), compared to a rescaled Gaussian with mean $\mu$ and standard deviation $\sigma$ (red dashed).}
\label{Fig_2}
\end{figure}

The reduced data cubes were analyzed as follows. From the calibrated data cubes we extracted spectra at the position of the magnetar SGR~1745$-$2900 (ON) using astrometry data \cite{10.1093/mnras/stx1439} and at several nearby positions chosen to sample the local background (OFF). See Fig. \ref{Fig_1} (top).  
The OFF spectra were averaged to form a reference $\langle$OFF$\rangle$, and the magnetar spectrum was differenced as DIFF $=$ ON $-$ $\langle$OFF$\rangle$.  
This subtraction removed large-scale GC emission and instrumental baselines while preserving narrow features at the source position. A high-pass filter was then applied to DIFF to suppress any residual continuum structure on $\gtrsim$1~GHz scales—Fig. \ref{Fig_1} (bottom). Within each SPW, a rolling median–absolute–deviation estimator was used to characterize the local noise per channel. These values were combined with a global root mean square (rms) to set adaptive clipping thresholds and construct per-channel weights, ensuring that regions affected by residual lines or artifacts contributed less to the analysis. Matched filtering was performed channel by channel using Gaussian kernels, with widths set by the expected axion-induced linewidth scaling with particle mass, $\Delta\nu/\nu \sim \Omega_* r_c \varepsilon^2/c$, where $r_c$ is the conversion altitude, $\varepsilon$ characterizes the eccentricity of the resonant-conversion geometry for an oblique rotator, and $c$ is the speed of light \cite{PhysRevD.102.023504, PhysRevLett.125.171301}. Each channel was fit with a weighted generalized least-squares regression that included a constant offset term, which accounts for any residual baseline level.  
For channels passing minimal coverage requirements—$>$70\% unmasked—, the fit returned an amplitude, its $1\sigma$ uncertainty, and the matched-filter signal-to-noise ratio (SNR).  
When the fitted amplitude was non-positive, a one-sided 95\% confidence upper limit was reported. 
Quality-control diagnostics confirmed that $\sim$80\% of all channels were retained after masking and filtering, with 163 frequency bins masked and ruled out for science—135.149–135.493\,GHz (45 bins), 136.480–136.738\,GHz (36), 146.828–147.250\,GHz (56), and 148.936–149.155\,GHz (26)—, and that the matched-filter SNR distribution was centered at zero with unit variance (mean $\mu \approx 0.02$; standard deviation $\sigma \approx 1.01$), consistent with Gaussian noise—Fig. \ref{Fig_2}.  
Median $1\sigma$ flux uncertainties per SPW were in the range $(3.5$–$4.6)\times10^{-5}$~Jy, corresponding to $\sim 40$–$50 \,\upmu$Jy/beam sensitivities when averaged over each SPW. See Table \ref{TableII}.  
No significant positive excesses in the SNR above $3.5\sigma$ were found, while the final product is a frequency-dependent upper-limit spectrum that can be used to constrain narrow axion-induced features in the SGR 1745--2900 environment.  

\section*{Results}
Goldreich \& Julian (GJ) developed a NS model with a corotating, charge-separated magnetospheric plasma characterized by the minimum charge density required to screen the rotation-induced electric field, in the idealized limit of perfect conductivity and a dipolar magnetic field \cite{1969ApJ...157..869G}. However, as shown by Beloborodov \cite{Beloborodov_2012, Beloborodov_2013}, the GJ model cannot satisfactorily explain the physical processes operating in magnetar magnetospheres, sustained by the creation of pair cascades and particle migration through magnetic forces, accelerating the charges to relativistic velocities, giving rise to curvature radiation, synchrotron emission and inverse Compton scattering \cite{1975ApJ...196...51R, 1979ApJ...231..854A, 1997A&A...322..846K}. In the magnetospheric regions where axions may resonate and produce millimeter waves within a Beloborodov-like model of SGR 1745–2900, the resonance altitude is expected to be a fraction of the stellar radius. Under these conditions, pair multiplicities—defined as the number of electron–positron pairs produced per primary particle over the GJ density, $\pazocal{M}=n_{e}/n_{_{\mathrm{GJ}}}$—are taken here to be characterized by the benchmark choice $\pazocal{M}\sim50$ and $\gamma_p\sim2$, interpreted as effective local parameters of the near-surface conversion region relevant to our mm-wave frequency range \cite{Beloborodov_2013}. The charge density of the magnetospheric plasma is $n_e = (2\Omega_* B/e)(\pazocal{M}/\gamma_p)$, with $\Omega_*=2\pi/P$ the angular velocity ($P$ the rotation period) and $e$ the elementary charge, leading to a plasma frequency $\omega_p^2 = 4\pi\alpha n_e/m_e$, with $\alpha$ the fine structure constant and $m_e$ the electron mass. Therefore, in order to interpret the observational results of the present manuscript, it is necessary to employ the mixing model capable of simultaneously accounting for a relativistic plasma and pair cascades in Ref. \cite{DEMIGUEL2025139328}. Following \cite{Beloborodov_2012, Beloborodov_2013}, simulations with SGR 1745--2900 are here performed in the limit of a weakly twisted dipolar field for a misaligned rotator. Let $\hat{m}$ be the magnetic axis, $\hat{z}$ the rotation axis, and $\hat{r}$ the photon propagation direction, $\theta_m$ the angle between the rotation and the magnetic axis with $f_1(t)=|3\,\mathrm{cos}\,\theta\,\hat{m}\cdot\hat{r}-\mathrm{cos}\,\theta_m|$ encoding the time dependence.  Resonant conversion occurs when $m_a\simeq\omega_p$, at a radius
\begin{equation}
r_c=224\,{\rm km}\, f_1^{1/3}(t)\,\frac{R_*}{10\,{\rm km}}
\left[\frac{B_0}{10^{14}\,{\rm G}}\frac{1\,{\rm s}}{P}
\left(\frac{1\,{\rm GHz}}{m_a}\right)^2\frac{\pazocal{M}}{\gamma_p}\right]^{1/3}\!,
\label{Eq.3}
\end{equation}
with $R_*$ the stellar radius. To estimate the expected flux density, we evaluate the resonant axion--photon conversion at the radius $r_c$ determined by the plasma condition above, following the approach in Ref. \cite{PhysRevLett.121.241102}. At that location, the magnetic field entering the mixing is the local dipolar field, $B(r_c)\propto B_0(R_*/r_c)^3$. The emitted signal is set by the flux of ambient DM axions crossing the resonant surface, which depends on the local density $\rho_a$, the infall velocity, and the effective area $\sim r_c^2$, together with the resonant axion--photon conversion in the magnetized pair plasma. Dividing the resulting radiated power by the geometric dilution factor $4\pi d^2$ and by the signal bandwidth gives the observed flux density. The axion velocity at the conversion point is $v_a^2\simeq 2GM_*/r_c$, with $G$ the gravitational constant and $M_*$ the stellar mass. The axion-induced line is broadened by various mechanisms \cite{PhysRevD.97.123001,PhysRevLett.121.241102,PhysRevD.102.023504,PhysRevLett.125.171301}; here the plasma–mirror scenario is adopted. The resulting flux density is \cite{DEMIGUEL2025139328}
\begin{equation}
\begin{aligned}
S_\nu = 1.6\times10^{-5}\,\upmu{\rm Jy}\, f_2(t) 
\left(\frac{1\,{\rm kpc}}{d}\right)^2
\left(\frac{g_{\gamma}}{10^{-12}\,{\rm GeV^{-1}}}\right)^2 \\
\times\frac{R_*}{10\,{\rm km}}
\left(\frac{m_a}{1\,{\rm GHz}}\right)^{4/3}
\left(\frac{B_0}{10^{14}\,{\rm G}}\right)^{1/3}
\left(\frac{\Omega_*}{1\,{\rm Hz}}\right)^{-8/3} \\
\times\frac{\rho_a}{0.4\,{\rm GeV\,cm^{-3}}}
\frac{M_*}{1 \,\mathrm{M}_\odot}
\frac{200\,{\rm km\,s^{-1}}}{v_0}
\frac{v_a}{c}
\left(\frac{\pazocal{M}}{\gamma_p}\right)^{-3/2}\,,
\end{aligned}
\label{Eq.4}
\end{equation}
where $f_2(t)=[3(\hat{m}\cdot\hat{r})^2+1]\,f_1^{-4/3}(t)$, $v_0$ is the radial velocity, and $\rho_a$ is the axion density. The telescope sensitivity in terms of coupling strength is given by Eq. \ref{Eq.5}.

\begin{figure*}[t]
\begin{equation}
\begin{split}
g_{\gamma} \gtrsim\;& 2.5\times10^{-10}\,{\rm GeV^{-1}}
\left(\frac{S_\nu}{1\,\upmu{\rm Jy}}\right)^{1/2}
f_2^{-1/2}(t)\,
\frac{d}{1\,{\rm kpc}}
\left(\frac{10\,{\rm km}}{R_*}\right)^{1/2}
\left(\frac{1\,{\rm GHz}}{m_a}\right)^{2/3}
\left(\frac{10^{14}\,{\rm G}}{B_0}\right)^{1/6}
\\
&\times
\left(\frac{\Omega_*}{1\,{\rm Hz}}\right)^{4/3}
\left(\frac{0.4\,{\rm GeV\,cm^{-3}}}{\rho_a}\right)^{1/2}
\left(\frac{1\,\mathrm{M}_\odot}{M_*}\right)^{1/2}
\left(\frac{v_0}{200\,{\rm km\,s^{-1}}}\right)^{1/2}
\left(\frac{c}{v_a}\right)^{1/2}
\left(\frac{\pazocal{M}}{\gamma_p}\right)^{3/4}.
\end{split}
\label{Eq.5}
\end{equation}
\end{figure*}

\begin{figure}[h]\centering
\includegraphics[width=.5\textwidth]{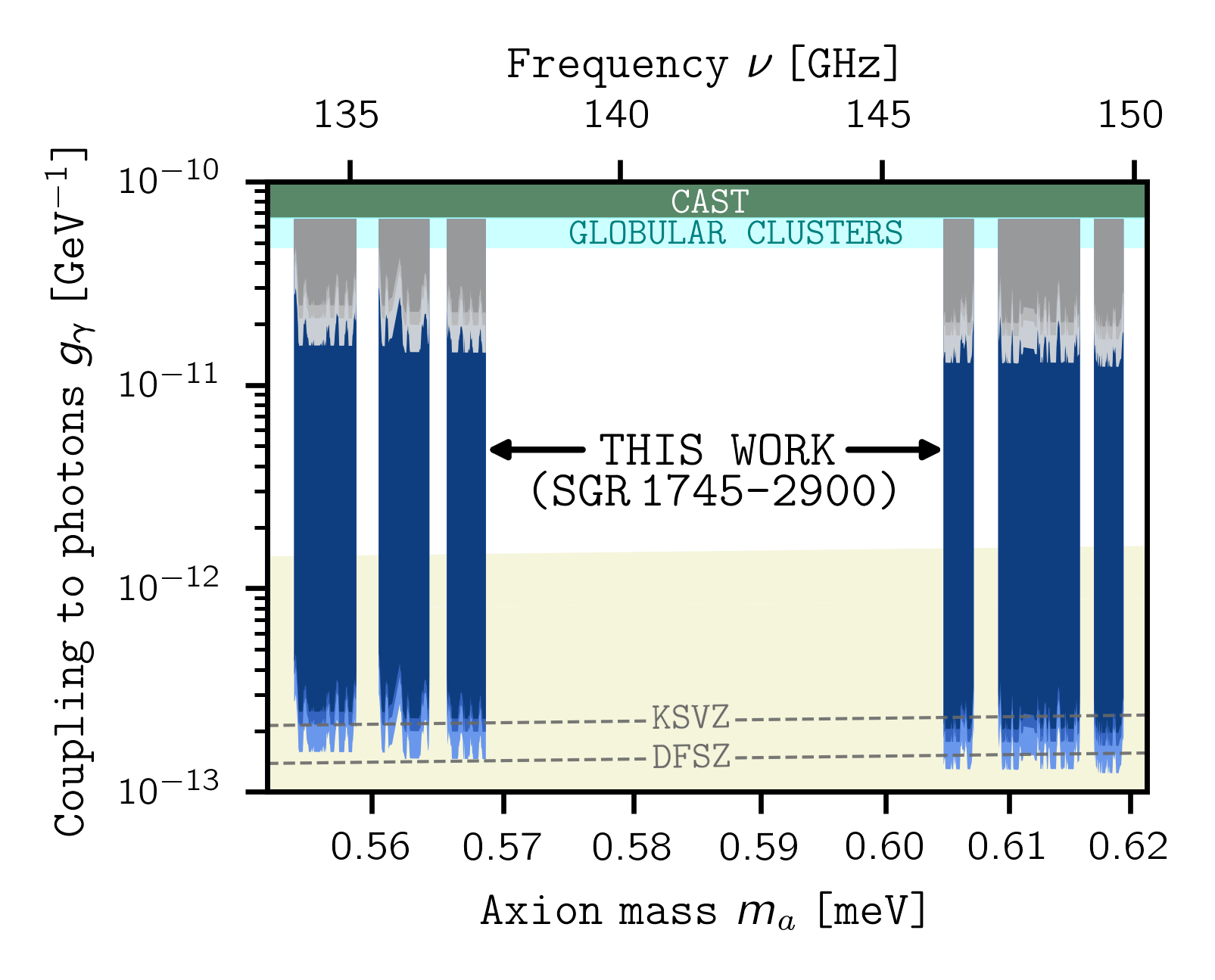}
\caption{95\% CL sensitivity to $g_{\gamma}$ from ALMA observations of SGR~1745$-$2900 for three NS configurations. The blue shaded regions correspond, from darker to lighter blue, to a \emph{light} star with (M$_*$, R$_*$, B$_0$) = (1.17 M$_\odot$, 12.54 km, 8.4 $\times$ 10$^{14}$ G), a \emph{canonical} NS with (1.40 M$_\odot$, 12.59 km, 9.52 $\times$ 10$^{14}$ G), and a \emph{heavy} model with (2.10 M$_\odot$, 12.13 km, 1.14 $\times$ 10$^{15}$ G), respectively, assuming a spike-enhanced DM density profile with $\rho_a = 6.4 \times 10^8$ GeV cm$^{-3}$ \cite{McMillan_2016,2018A&A...619A..46L}, which we take as an optimistic scenario. The gray shaded regions show the corresponding projections for the reference NFW profile adopted in this work, again ordered from darker to lighter shading for the light, canonical, and heavy stellar models, respectively, corresponding to a more conservative DM assumption in which no central spike around the SMBH is included. For all configurations we adopt $P = 3.76$ s \cite{Mori:2013yda}, $v_0 = 236$ km s$^{-1}$ \cite{Bower:2014tea}, $\theta \simeq 20^\circ$, $\theta_m \simeq 10^\circ$, $\varepsilon \approx 0.9$, and a Beloborodov-like magnetosphere, with effective local plasma parameters $\pazocal{M} \simeq 50$ and $\gamma_p \simeq 2$ in the resonant region \cite{Beloborodov_2012, Beloborodov_2013}. Results are compared with CAST limits \cite{PhysRevLett.133.221005, CAST:2017uph} and globular-cluster bounds \cite{Ayala:2014pea, 2022JCAP...10..096D}. The \emph{phenomenologically preferred} axion window is shown in beige, together with benchmark QCD-axion models (KSVZ, DFSZ) \cite{DiLuzio:2016sbl, PhysRevLett.43.103, Shifman1980CanCE, DINE1981199, osti_7063072}.}
\label{Fig_3}
\end{figure}

From the upper--limit amplitudes derived in the matched--filter analysis, we computed the benchmark limits on the axion--photon coupling strength, $g_{\gamma}$, using Eq. \ref{Eq.5}. Each spectral channel was mapped to an axion mass through the resonance condition, with the conversion radius, plasma multiplicity, and Lorentz factor obtained from magnetospheric profiles adapted to Beloborodov-type models, within a resonant-scattering picture successfully applied to magnetars' soft X-ray emission \cite{Rea:2008zs}. We adopt this framework as a concrete benchmark for the near-surface plasma conditions relevant to resonant conversion, while noting that alternative magnetospheric descriptions have also been discussed in the literature and may lead to different projections for $g_{\gamma}$—see, e.g., Refs.~\cite{2002ApJ...574..332T, 10.1111/j.1365-2966.2008.13125.x, Thompson_2020}. The axion velocity at the resonance region was included via gravitational acceleration, while geometric and temporal factors were treated explicitly. In particular, the time--dependent factor $f_{2}(t)$ was averaged over one stellar rotation to account for the multi--hour integration of our ALMA observations. For every contiguous frequency run, we computed the one-sided 95\% confidence level upper limits $g_{\gamma}^{95}$, which are summarized in Table~\ref{TableIII}. In Fig. \ref{Fig_3}, we adopt the NFW profile modified to include a spike (sp) with radial index $\gamma_{\mathrm{sp}} = 7/3$ and radius $R_{\mathrm{sp}} = 0.1~\mathrm{kpc}$, consistent with the 99.7\% upper limit on deviations from a purely black-hole orbit of the S2 star around Sgr~A$^\ast$, following Refs.~\cite{McMillan_2016, 2018A&A...619A..46L, Darling:2020uyo, Darling:2020plz}. This model yields a DM density of $\rho_{\mathrm{sp}} \sim 6.4 \times 10^{8}~\mathrm{GeV\,cm^{-3}}$ at the position of the magnetar. We employ three benchmark models (\textit{light}, \textit{canonical}, \textit{heavy}) for the stellar structure, derived self-consistently by solving the Tolman-Oppenheimer-Volkoff (TOV) equation \cite{PhysRev.55.374} using the BSk24 nuclear equation of state (EoS) \cite{Mutafchieva_2020} to produce the mass--radius--moment-of-inertia relations. This EoS has been chosen among the many possibilities as it is a modern option that proved capable of accounting for recent NS cooling observations \cite{Marino:2024gpm}. Namely, for a given value of the central pressure, we obtain from the TOV the gravitational mass $M_*$, radius $R(M_*)$ and moment of inertia $I(M_*)$. In turn, the moment of inertia determines the dipolar magnetic field from the timing solution through the spin-down expression $B_{0} \simeq 6.4\times10^{19}\,I_{45}^{1/2}\,\bigl(P\,\dot{P}\bigr)^{1/2}\ {\rm G}$, where $I_{45}\equiv I/10^{45}\,{\rm g\,cm^{2}}$, $P$ is measured in seconds and $\dot{P}$ in s\,s$^{-1}$. The observed NS population spans masses from about $1.2\,M_\odot$ \cite{Martinez_2015} to $2.1\,M_\odot$ \cite{2020NatAs...4...72C}, and within this physical range we consider three benchmark configurations: a light model with $M_*=1.17\,M_\odot$, yielding $R=12.54\,$km, $I_{45}=1.55$, and $B_{0}=8.4\times10^{14}\,$G; a canonical model with $M_*=1.40\,M_\odot$, yielding $R=12.59\,$km, $I_{45}=1.99$, and $B_{0}=9.52\times10^{14}\,$G; and a heavy model with $M_*=2.1\,M_\odot$, yielding $R=12.13\,$km, $I_{45}=2.78$, and $B_{0}=1.14\times10^{15}\,$G.
\begin{table}[h]
\centering
\caption{One-sided 95\% CL limits on $g_{\gamma}$ per SPW for the \textit{canonical} NS model (see Fig.~\ref{Fig_3}), shown for two DM profile assumptions at the position of SGR~1745--2900: an NFW profile with $\rho_{\rm NFW}=6.5\times10^{4}\,\mathrm{GeV\,cm^{-3}}$ \cite{Darling:2020plz} and a DM spike with $\rho_{\rm sp}=6.4\times10^{8}\,\mathrm{GeV\,cm^{-3}}$ \cite{McMillan_2016,2018A&A...619A..46L}. For each SPW we report the minimum, mean, and median channel-wise $g_{\gamma}^{95}$ values.}
\begin{tabular}{c ccc ccc}
\hline\hline
& \multicolumn{3}{c}{NFW profile} & \multicolumn{3}{c}{DM spike} \\
\cline{2-4} \cline{5-7}
SPW & min. & mean & median & min. & mean & median \\
& \multicolumn{3}{c}{$g_{\gamma}^{95}$ [$10^{-11}$ GeV$^{-1}$]} & \multicolumn{3}{c}{$g_{\gamma}^{95}$ [$10^{-13}$ GeV$^{-1}$]} \\
\hline
1 & 2.11 & 2.46 & 2.25 & 2.13 & 2.48 & 2.27 \\
2 & 1.96 & 2.25 & 2.14 & 1.98 & 2.27 & 2.16 \\
3 & 1.74 & 1.96 & 1.80 & 1.75 & 1.98 & 1.81 \\
4 & 1.68 & 1.93 & 1.82 & 1.69 & 1.95 & 1.83 \\
\hline\hline
\end{tabular}
\label{TableIII}
\end{table}

\section*{Conclusions}
SGR~1745$-$2900 is one of the most promising astrophysical targets to probe axion DM, owing to its extreme surface magnetic field and the plausible enhancement of the DM density in the vicinity of Sgr~A* at the Galactic Center. Since the resonance condition for axion--photon conversion depends on the ratio $\pazocal{M}/\gamma_p$---where $\pazocal{M}$ denotes the pair multiplicity and $\gamma_p$ the plasma Lorentz factor---values $\pazocal{M}/\gamma_p \sim 1$ favor efficient conversion at microwave frequencies, whereas $\pazocal{M}/\gamma_p \gg 1$ shift the accessible range from microwaves into the millimeter regime. The present work focuses on the latter case.

We have carried out the first mm-wave search for axion DM using 4.8~h of ALMA observations of SGR~1745$-$2900, finding no statistically significant evidence for axion-induced narrow emission lines. Interpreted within a Beloborodov-like magnetospheric framework with resonant axion--photon conversion, this null result can be translated into exclusion limits in the sub- to near-meV mass range, specifically over the intervals 0.554--0.562, 0.562--0.570, 0.604--0.611, and 0.612--0.620~meV, as shown in Fig.~\ref{Fig_3}.

Within our set of stellar benchmarks, we estimate that the minimum DM density required at the position of the magnetar for our exclusion to remain sensitive to the preferred QCD axion window is of order $\rho_a \sim (3$--$6)\times10^7~{\rm GeV\,cm^{-3}}$. In an NFW$+$spike description, this could be achieved by a spike somewhat milder than our fiducial benchmark $(\gamma_{\rm sp}=7/3,\;R_{\rm sp}=0.1~{\rm kpc})$, for example with $\gamma_{\rm sp}\approx 1.9$--$2.0$ at fixed $R_{\rm sp}=0.1~{\rm kpc}$, or alternatively by a more compact spike with $\gamma_{\rm sp}=7/3$ and $R_{\rm sp}\sim 10$--$20~{\rm pc}$. By contrast, the more compact Gondolo--Silk~\cite{PhysRevLett.83.1719} benchmark considered in Ref.~\cite{laTorreLuquePedro:2024est}, with the same canonical inner slope $\gamma_{\rm sp}=7/3$ but a smaller characteristic radius $R_{\rm sp}\simeq 0.34~{\rm pc}$, yields a DM density at the position of SGR~1745$-$2900 of only $\rho_{\rm sp}\approx 3.7\times10^5~{\rm GeV\,cm^{-3}}$. Within our Beloborodov-like magnetospheric framework, this would weaken the inferred sensitivity to approximately $g_\gamma^{95}\sim (6$--$10)\times10^{-12}~{\rm GeV^{-1}}$ across the range spanned by our light and heavy stellar benchmarks. More generally, adopting a cored inner profile would further reduce the DM density at the location of SGR~1745$-$2900 and hence further relax the inferred limits.

The detection sensitivity scales with the magnetospheric plasma properties mainly as $g_\gamma \propto (\pazocal{M}/\gamma_p)^{3/4}$. Accordingly, plausible variations in the plasma conditions within Beloborodov-like twisted-outflow models are expected to act primarily as a rescaling of the coupling reach. Changes in basic NS parameters such as mass and radius can likewise shift the sensitivity, typically by a factor of a few or less. In this sense, the values adopted in Table \ref{TableIII} and Fig.~\ref{Fig_3} should be regarded—for our light, canonical, and heavy NS configurations—as effective local parameters of the near-surface resonant conversion region rather than as global descriptors of the full magnetosphere. Over the portion of the outflow most relevant for resonant conversion in our frequency range, the plasma conditions are expected to remain comparable at the order-of-magnitude level, so benchmark values such as $\pazocal{M}\sim 50$ and $\gamma_p\sim 2$ may be taken as representative of the conversion zone in an effective sense. Alternative magnetospheric descriptions may nevertheless differ in plasma stratification, charge supply, current closure, and velocity structure, thereby modifying the local plasma-frequency profile, the resonant radius, and hence the overall normalization of the corresponding limits. For the present ALMA data set, a comparison with a recent alternative realization suggests that the residual impact on the inferred $g_{\gamma}$ sensitivity may remain at the level of order-unity factors; see the Note added.

Our approach provides access to a region of parameter space that remains difficult to probe by other methods. The search presented here is therefore complementary to next-generation laboratory experiments~\cite{DeMiguel:2020rpn, PhysRevD.109.062002, sym16020163, Hernandez-Cabrera:2023syh, PhysRevD.111.023016}, and can be extended toward improved sensitivity and broader frequency coverage through the same methodology and through parallel experimental strategies.

\paragraph*{Note added}
We also note the recent simulations in Ref.~\cite{Roy:2025mqw}, which are based on an alternative framework to that of Ref.~\cite{DEMIGUEL2025139328}. The projected $g_{\gamma}$ sensitivities in Ref.~\cite{Roy:2025mqw} are derived within an alternative magnetospheric description based on Kostenko \& Thompson \cite{Thompson_2020}, and use a different DM profile together with benchmark NS parameters distinct from those adopted in the present Letter. After accounting for these differences, the corresponding ALMA sensitivities in the near-meV range become broadly comparable to those inferred here from observations within a Beloborodov-like treatment, with the remaining differences lying at the level of factors of order unity, and in the most representative cases at the few-tens-percent level.

\section*{Acknowledgements}
This paper makes use of the following ALMA data:\newline ADS/JAO.ALMA\#2024.1.00310. The project that gave rise to these results received the support of a fellowship from “la Caixa” Foundation (ID 100010434). The fellowship code is LCF/BQ/PI24/12040023. J.DM. acknowledges support from the Spanish Ministry of Science, Innovation and Universities and the Agency (EUR2024-153552 financed by MICIU/AEI/10.13039/501100011033). E.H. has received funding from the European Union’s Horizon Europe research and innovation program under grant agreement No. 101188037 (AtLAST2). F.P. acknowledges support from the MICINN under grant numbers PID2022-141915NB-C21. N.R. is supported by the ERC via the Consolidator grant “MAGNESIA” (No. 817661), the ERC Proof of Concept ”DeepSpacePULSE” (No. 101189496), the Spanish grant PID2023-153099NA-I00, and by the program Unidad de Excelencia María de Maeztu CEX2020-001058-M. D.D.G. is supported by a Juan de la Cierva fellowship (JDC2023-052264-I). Thanks L. Maud, H. Messias, and A. Richards for support.

\bibliographystyle{elsarticle-num}
\bibliography{sample631}






\end{document}